\documentclass[11pt, a4paper]{article}

\usepackage[utf8]{inputenc} 
\usepackage[T1]{fontenc}    
\usepackage{hyperref}       
\usepackage{url}            
\usepackage{booktabs}       
\usepackage{amsfonts}       
\usepackage{nicefrac}       
\usepackage{microtype}      
\usepackage{xcolor}         
\usepackage{bm}             
\usepackage[ruled,vlined,linesnumbered,lined,boxed]{algorithm2e}
\usepackage{multirow}
\usepackage{tikz}
\usepackage{pgfmath}
\usepackage{subcaption}
\usepackage{pgfplots}
\usepackage[T1]{fontenc}
\usepackage{lmodern}
\usepackage[numbers,sort&compress]{natbib}
\usetikzlibrary{calc,intersections}

\title{Cellular Learning: Scattered Data Regression in High Dimensions via Voronoi Cells}

\author{%
  Shankar P.~Sastry\\
  Somerville, MA 02145 \\
  \texttt{shankar.prasad@gmail.com}
  \thanks{The author would like to thanks Ms. Christine Pickett for reviewing a draft of the paper and suggesting numerous changes to improve the quality of the paper.}
}

\date{\today}

\begin{document}
\maketitle
\begin{abstract}
  I present a regression algorithm that provides a continuous, piecewise-smooth function approximating scattered data. It is based on composing and blending linear functions over Voronoi cells, and it scales to high dimensions. The algorithm infers Voronoi cells from seed vertices and constructs a linear function for the input data in and around each cell. As the algorithm does not explicitly compute the Voronoi diagram, it avoids the curse of dimensionality. An accuracy of around 98.2\% on the MNIST dataset with 722,200 degrees of freedom (without data augmentation, convolution, or other geometric operators) demonstrates the applicability and scalability of the algorithm.
\end{abstract}
\section{Introduction}
I introduce an algorithm to perform regression of scattered data. The algorithm scales to high dimensions, and I demonstrate it by classifying the images from the Modified National Institute of Standards and Technology (MNIST) dataset~\cite{deng2012mnist} with an accuracy that is comparable to state-of-the-art techniques when no data augmentation or geometric operators are used. Many machine learning algorithms approximate scattered data with suitable computational models. Even classification algorithms treat input data as probabilities and build a model that computes the probability of a new data point belonging to one of the classes. These algorithms differ in their ability to model complex relationships, to scale to large input data, to scale to high dimensions, to explain/interpret the results, etc. 

The linear regression algorithm scales well with the number of data points and dimensions~\cite{bishop05,murphy13}. It can also be easily interpreted. It, however, cannot handle nonlinear data. The logistic regression algorithm has the same attributes, too~\cite{bishop05,murphy13}. Some algorithms compute a continuous, piecewise-linear function approximating input data~\cite{toriello12}, but they do not scale to high dimensions because they work on regular grids, whose complexity is exponential in the number of dimensions. Siahkamari et al.~\cite{siahkamari20} provide an algorithm to return a continuous, piecewise-linear function that is represented as a difference of two continuous, piecewise-linear convex functions. The convex functions are constructed using a max-affine regression model, where the function is computed as a maximum over many affine functions. The algorithm scales quadratically with the dimensionality of the data, and the authors have validated its scalability for problems involving up to 15 dimensions. The support vector machine (SVM) and support vector regression (SVR) algorithms can handle nonlinear data with nonlinear kernels~\cite{cortes-vapnik-ml95}, and they are somewhat interpretable~\cite{vanbelle2016}. They provide a smooth function as the output, but they are computationally expensive due to quadratic programming techniques involved in training the model. They can be accelerated by finding approximate solutions to the quadratic programming problem~\cite{tsang05,wang17}. Neural networks with multiple layers are not interpretable, but they scale well with stochastic optimization techniques~\cite{kingma14}. With the rectified linear unit (ReLU) activation function, neural networks output a continuous, piecewise-linear function. The activation function has been extensively used in practice~\cite{fukushima80,chen22}. Radial basis function (RBF) networks can also handle nonlinear data, and they output a smooth function~\cite{Broomhead1988a, Broomhead1988b}. They are also interpretable as only a single hidden layer is present, but just as SVM and SVR, they are computationally expensive to train. 

The observations above indicate that linear functions are easier to compute compared to nonlinear functions. They are also easier to interpret. On the other hand, multiple layers in neural networks make the models hard to interpret. Driven by these observations, I developed an algorithm that computes multiple linear functions approximating the data within and around a set of seed vertices such that each seed vertex has a corresponding linear function. The functions are then blended to construct a piecewise smooth function that approximates the scattered input data. The location of the seed vertices, the coefficients of the monomials in the linear function, and the extent of the blending are all parameters in the model that are determined using a stochastic optimization algorithm. The technique that blends the linear functions is the main contribution of this paper. The technique does not explicitly compute the Voronoi diagram of the seed vertices. It evades the curse of dimensionality by focusing on a single Voronoi cell or a pair of cells, but not multiple cells together. I call this algorithm the cellular learning algorithm.

There are a couple of techniques that resemble the cellular learning algorithm. One such technique is the Voronoi boundary classification technique by Polianskii and Pokorny~\cite{polianskii19}. They use a Monte-Carlo-based approach to compute a weighted integral over the boundaries of Voronoi cells to determine the classification. They do not compute the exact integral because it is impractical at high dimensions. They have demonstrated that their algorithm scales to high dimensions with their implementation. They published another technique~\cite{polianskii20}, where they used a randomized approximation approach to compute Voronoi diagrams in high dimensions to avoid the curse of dimensionality. In the cellular learning algorithm, there are no approximations of the Voronoi diagram or randomization.

RBF networks~\cite{Broomhead1988a, Broomhead1988b} also resemble the cellular learning algorithm. RBF networks use a fixed set of seed vertices that are computed using the $k$-means algorithm~\cite{lloyd82}. They use RBFs that are compactly supported, i.e., the function value tends to $0$ as the point at which the function is evaluated moves away from a seed vertex, but it is generally never equal to $0$. The weights of RBFs in the network are the only parameters that are computed via a numerical optimization algorithm. The cellular learning algorithm differs from this technique in a few ways. First, the basis function it uses is not radial. The value of the basis function is $1$ within the Voronoi cell of the seed vertex, and its value becomes $0$ at a finite distance from the seed vertex. Second, each Voronoi cell of a seed vertex has an associated linear function instead of a constant. RBF networks can also be extended in this way, but to my knowledge, it has not been studied much. Third, the location of the seed vertices is also determined via a numerical optimization algorithm after its initialization.

The cellular learning algorithm is designed to reduce the errors due to both bias and variance. The algorithm uses more seed vertices to mitigate high bias. In order to mitigate high variance, an L2 regularization technique is used along with the regularization of the parameters associated with the blending of functions. In combination, the algorithm provides good accuracy for the high-dimensional MNIST data set against which it was tested.

\section{Background}
The cellular learning algorithm uses the $k$-Means algorithm~\cite{lloyd82} to initialize the seeds around which Voronoi cells~\cite{voronoi1908a} are implicitly constructed. It uses the Adam optimization algorithm~\cite{kingma14} to move the seeds and determine a linear function that approximates the data around the seed. I provide a brief background on these techniques in this section. I refer to a specified set of named points as \emph{vertices}. A \emph{point} may refer to any point in space.
\subsection{The Lloyd's $k$-Means Clustering Algorithm}
The Lloyd's $k$-Means clustering algorithm~\cite{lloyd82} is an unsupervised machine learning algorithm, where $k$ seed vertices are randomly assigned among the input scattered data of $n$ observation vertices. The observation vertices are then distributed among the seeds to form $k$ clusters. Each observation vertex is assigned to its closest seed, and the set of observation vertices with the same seed vertex form a cluster. The seed vertex locations are then updated to the centroid of their respective cluster of assigned observation vertices. This process is iteratively repeated until the clusters have converged. 

\subsection{Voronoi Diagram}

\begin{figure}
\centering
\includegraphics[width=0.25\linewidth]{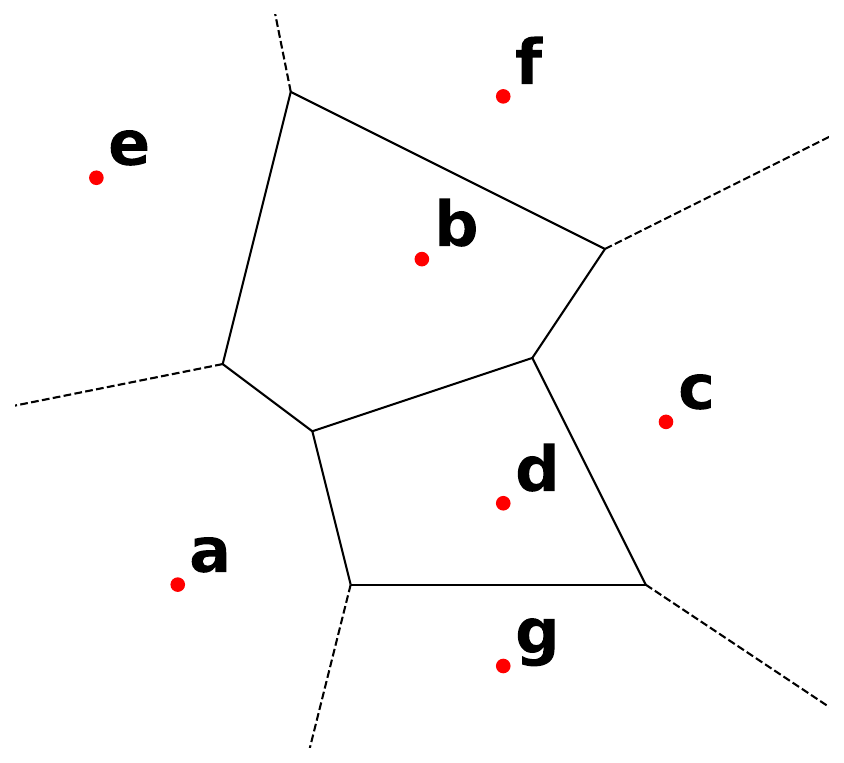}
\caption{An example of a Voronoi diagram in 2D for a given set of vertices. Note that every line segment or ray (dashed lines) in the diagram is a segment from the perpendicular bisector of a pair of vertices.}
\label{fig:voronoi}
\end{figure}

A Voronoi diagram~\cite{voronoi1908a} is a partition of a space into regions close to each given seed vertex. The region around a seed vertex is such that for all points in the region, its own seed vertex is the closest seed vertex among all seed vertices. A Voronoi cell is the region around a given seed vertex. The relationship between the Voronoi diagram and a $k$-mean algorithm-derived clusters is that the boundaries of clusters constitute the Voronoi diagram of seed vertices. Given two seed vertices in two dimensions, the Voronoi diagram is simply the perpendicular bisector of the line joining the two vertices, and the Voronoi cells are halfplanes of either side of the perpendicular bisector. In two dimensions, when multiple vertices are provided as the input, the Voronoi diagram consists of line segments from perpendicular bisectors of some pairs of input vertices as shown in Fig.~\ref{fig:voronoi}. In higher dimensions, the Voronoi diagram consists of hyperpolygons on the perpendicular hyperplanes that bisect the line joining a pair of input vertices.

\subsection{Adam Optimization Algorithm}
The Adam optimization algorithm~\cite{kingma14}, short for adaptive moment estimation, is an extension of the stochastic gradient descent algorithm that adapts the learning rate for each parameter based on the history of gradients. The algorithm is known to work well for neural networks, and I used this algorithm to estimate the optimal location of seed vertices, the parameters that define the linear function for each cell, and the blending parameter for every cell.

\section{Cellular Learning}
The cellular learning algorithm constructs a network of Voronoi cells that define the function that approximates the input scattered data. I call the network of cells a cellular network. Each cell has a set of parameters to define the function. As in any machine learning algorithm, one has to ensure neither high bias nor high variance results in a suboptimal approximation. An excessive number of parameters may reduce the bias by preventing underfitting of the data, but it can increase the variance due to overfitting. Thus, I have also introduced hyperparameters designed to prevent overfitting. All parameters should be fine-tuned to obtain an optimal approximation. This section describes the algorithm in detail.

\subsection{Input to the Algorithm} \label{subsection:input}
The algorithm takes scattered data in any dimension as its input. Let $(\bm{a_0}, b_0)$, $(\bm{a_1}, b_1)$, ..., $(\bm{a_{n-1}}, b_{n-1})$ be the input dataset, where $\bm{a_i} \in \mathbb{R}^d$ ($0 \le i < n$) is a $d$-dimensional vector and $b_i$ ($0 \le i < n$) is a scalar. The dataset may be sourced from an unknown function $f(\bm{x})$ and/or be noisy. The algorithm computes an approximate function $\hat{f}(\bm{x})$, where $\bm{x} \in \mathbb{R}^d$ is a $d$-dimensional vector and $\hat{f}(\bm{a_i}) \approx b_i$ for $0 \le i < n$.

\subsection{Output of the Algorithm: Cellular Network} \label{subsection:parameters}
A cellular network is defined using a few hyperparameters and parameters.
The first hyperparameter that defines a cellular network is the number of cells in it. Let this number be $k$. Each cell constitutes a linear function in $d$ dimensions. Let the coordinate axes in $d$ dimensions be represented by $x_j$, where $1 \le j \le d$. The linear function for cell $i$ ($0 \le i < k$) is defined as
\begin{equation}
L_i = \beta_{i0} + \sum_{j = 1}^{d} \beta_{ij} x_j = \bm{\beta_i}\cdot\bm{x},
\label{eq:linear_function}
\end{equation}
where $0 \le i < k$, $\bm{\beta_i} = [\beta_{i0}, \beta_{i1}, ... \beta_{id}]  \in \mathbb{R}^{d + 1}$ is a $d + 1$-dimensional vector, and $\bm{x} = [1, x_1, x_2, ... x_d] \in \mathbb{R}^{d + 1}$ is also a $d + 1$-dimensional vector. The vectors $\bm{\beta_i}$ ($0 \le i < k$) are parameters used to define the cellular network. The location of the vertices (also called seeds or sites) that define the Voronoi cells are also parameters that define a cellular network. I will denote them as $\bm{c_i}$, where $0 \le i < k$ and $\bm{c_j}  \in \mathbb{R}^d$ is a $d$-dimensional vector. Additionally, every cell also uses a scalar \emph{blending} parameter, $\alpha_i$ ($0 \le i < k$), which controls the extent to which the linear function blends into neighboring space. I will formally describe how the blending parameter is used in Section~\ref{subsection:weight} below. If all the parameters in this section are provided, the value of the approximate function can be computed anywhere. There are other hyperparameters used for regularization, which I will describe in ~\ref{subsection:regularization}. Those hyperparameters are needed only to prevent overfitting and not to compute the function value at any point.

Note that the total number of parameters in a cellular network with $k$ seed vertices in $d$ dimensions is $k * 2 * (d + 1)$. For a cell $i$, the linear function needs $d + 1$ parameters for the vector $\bm{\beta_i}$, $d$ parameters for the vector $\bm{c_i}$, and the blending parameter, $\alpha_i$, which counts as one additional parameter.

\subsection{Weight Computation and Blending} \label{subsection:weight}
In this section, I will assume that the hyperparameters and parameters mentioned above are known to us. I will describe how to compute the approximate function $\hat{f}(\bm{x})$ at any point $\bm{x}$ given the hyperparameters and parameters. As mentioned above, every cell in a cellular network has a linear function. These linear functions have to be weighted appropriately over the whole domain such that the resulting function is continuous. I will first treat each linear function independently and describe how its weight varies over the domain. I will then describe how the normalization of all the weights from all linear functions in the network blends the linear functions to obtain a continuous, piecewise-smooth function. 

Assume that the cellular network has $k$ cells implicitly defined using $k$ vertices. Let the linear function associated with cell $i$ be $L_i$. For any point $\bm{p}$,
\begin{equation}
\hat{f}(\bm{p}) = \sum_{i = 0}^{k - 1} \omega_i(\bm{p}) L_i(\bm{p}),
\label{eq:weighted_linear_function}
\end{equation}
where $\hat{f}(\cdot)$ is the approximate function computed by the cellular learning algorithm, and $\omega_i(\bm{p})$ is the weight of the linear function $L_i$ at $\bm{p}$. This section focuses on computing $\omega_i(\bm{p})$ for $0 \le i < k$ at any $\bm{p}$ in the domain of the function. 

In order to compute $\omega_i(\bm{p})$ for a given $i$, we have to first compute the relative weights $\omega_i^{\mathrm{rel}}(\bm{p})$ for all $0 \le i < k$ and then normalize the weights so that they sum to $1$. The relative weight is $1$ if $\bm{p}$ is on or inside the Voronoi cell of vertex $\bm{c_i}$. One can easily check the distance from $\bm{p}$ to $\bm{c_i}$ for $0 \le i < k$ and find the closest one(s). The relative weights for the linear function corresponding to the closest vertex (vertices) is (are) set to $1$.

In order to obtain a continuous function ($\hat{f}(\cdot)$), if $\bm{p}$ is outside the Voronoi cell, the relative weight, $\omega_i^{\mathrm{rel}}(\bm{p})$, should continuously reduce to $0$ as $\bm{p}$ moves away from the boundary of the Voronoi cell, i.e., the weight should be $1$ on the Voronoi cell boundary and slowly reduce to $0$ at some distance. The distance at which $\omega_i^{\mathrm{rel}}(\bm{p})$ vanishes is controlled by the blending parameter, $\alpha_i$ ($0 \le i < k$), referenced in Section~\ref{subsection:parameters}. Clearly, some distance (the closest distance or otherwise) from $\bm{p}$ to the boundary of the Voronoi cell of $\bm{c_i}$ should be computed to find the relative weight.

If one considers the closest distance to the boundary of the Voronoi cell of $\bm{c_i}$ from any point $\bm{p}$, one would essentially reconstruct the Voronoi diagram. The complexity of computing the Voronoi diagram for $n$ vertices is $\Theta(n^{\lceil \frac{d}{2} \rceil})$, where $d$ is the number of dimensions. Clearly, this approach is prohibitive for high-dimensional data.

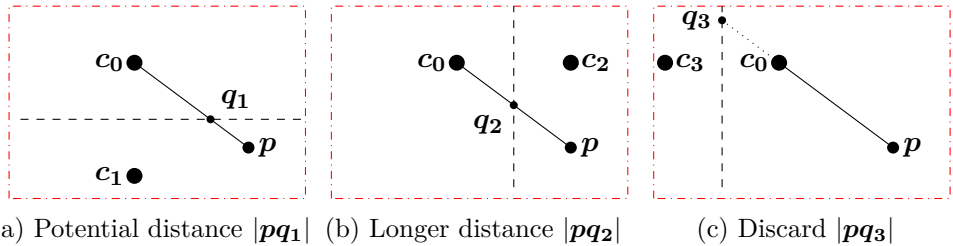
\begin{figure}
\centering
    \subfloat[Potential distance $\left|\bm{pq_1}\right|$]{
        \begin{tikzpicture}[scale=1.5]
          \def\xmin{-1.1}
          \def\xmax{1.5}
          \def\ymin{-0.2}
          \def\ymax{1.5}

          \draw[dashdotted, red] (\xmin, \ymin) rectangle (\xmax, \ymax);

		  \coordinate (A) at (0, 1);
		  \coordinate (B) at (0, 0);
		  \coordinate (P) at (1, 0.25);
		  \coordinate (I) at (0.667, 0.5);

		  \fill (A) circle (2pt) node[left] {$\bm{c_0}$};
		  \fill (B) circle (2pt) node[left] {$\bm{c_1}$};
		  \fill (P) circle (1.5pt) node[right] {$\bm{p}$};
		  \fill (I) circle (1pt) node[anchor=south west] {$\bm{q_1}$};

		  \draw[dashed] (-1, 0.5) -- (1.5, 0.5); 
		  \draw (A) -- (P); 
		\end{tikzpicture}
    }\hfill
    \subfloat[Longer distance $\left|\bm{pq_2}\right|$]{
        \begin{tikzpicture}[scale=1.5]
          \def\xmin{-1.1}
          \def\xmax{1.5}
          \def\ymin{-0.2}
          \def\ymax{1.5}

          \draw[dashdotted, red] (\xmin, \ymin) rectangle (\xmax, \ymax);
          
		  \coordinate (A) at (0, 1);
		  \coordinate (C) at (1, 1);
		  \coordinate (P) at (1, 0.25);
		  \coordinate (I) at (0.5, 0.625);

		  \fill (A) circle (2pt) node[left] {$\bm{c_0}$};
		  \fill (C) circle (2pt) node[right] {$\bm{c_2}$};
		  \fill (P) circle (1.5pt) node[right] {$\bm{p}$};
		  \fill (I) circle (1pt) node[anchor=north east] {$\bm{q_2}$};

		  \draw[dashed] (0.5, -0.1) -- (0.5, 1.5); 
		  \draw (A) -- (P); 
		\end{tikzpicture}
    }\hfill
    \subfloat[Discard $\left|\bm{pq_3}\right|$]{
        \begin{tikzpicture}[scale=1.5]
          \def\xmin{-1.1}
          \def\xmax{1.5}
          \def\ymin{-0.2}
          \def\ymax{1.5}

          \draw[dashdotted, red] (\xmin, \ymin) rectangle (\xmax, \ymax);
          
		  \coordinate (A) at (0, 1);
		  \coordinate (D) at (-1, 1);
		  \coordinate (P) at (1, 0.25);
		  \coordinate (I) at (-0.5, 1.375);

		  \fill (A) circle (2pt) node[left] {$\bm{c_0}$};
		  \fill (D) circle (2pt) node[right] {$\bm{c_3}$};
		  \fill (P) circle (1.5pt) node[right] {$\bm{p}$};
		  \fill (I) circle (1pt) node[left] {$\bm{q_3}$};

		  \draw[dashed] (-0.5, -0.1) -- (-0.5, 1.5); 
		  \draw (A) -- (P); 
		  \draw[dotted] (A) -- (I); 
		\end{tikzpicture}
    }
    \caption{Given a point $\bm{p}$, how can one find its distance from the boundary of the Voronoi cell of vertex $\bm{c_0}$? The solution is to consider other vertices ($\bm{c_1}$, $\bm{c_2}$, and $\bm{c_3}$), and find the hyperplanes that perpendicularly bisect $\bm{c_0}$ and the other vertex. In the figures above, the dashed lines are the hyperplanes. We find the distance from $\bm{p}$ to the hyperplane along the line joining $\bm{p}$ and $\bm{c_0}$. This distance can be found by solving a linear equation. We should find the shortest distance from $c_0$ for which the point of intersection is on line segment $\bm{pc_0}$. The distances in (a) and (b) are both considered as they are on the line segment. In (c), the point $\bm{q_3}$ is not on the line segment $\bm{pc_0}$, so it is discarded. Since $\left|\bm{c_0q_2}\right| < \left|\bm{c_0q_1}\right|$, $\left|\bm{pq_2}\right|$ is considered the distance from $\bm{p}$ to the boundary of the Voronoi cell of $\bm{c_0}$. }
    \label{fig:distances}
\end{figure}

Instead of finding the closest distance to the boundary of the Voronoi cell, I compute the distance from $\bm{p}$ to the boundary of the Voronoi cell along the line joining $\bm{p}$ to $\bm{c_i}$. The complexity of computing this distance for a point $\bm{p}$ for all $n$ cells in a $d$-dimensional space is $\Theta(dn^2)$, but it can be reduced to $\Theta(d\log{(n)})$ using a hierarchical approach (briefly described in Section ~\ref{section:future}), where the distance is only computed among $\Theta(\log{n})$ vertices, but that is a subject of future research. I will explain the algorithm with $\Theta(dn^2)$ time complexity below. See Fig.~\ref{fig:distances} for more description.

Consider a point $\bm{p}$ that is not on or inside the Voronoi cell of vertex $\bm{c_i}$. The boundary of the Voronoi cell of $c_i$ intersects the line joining $\bm{p}$ and $\bm{c_i}$ at a point $\bm{q}$ on the hyperplane that is perpendicular to the line joining $\bm{c_i}$ and some other vertex $\bm{c_j}$ and contains the midpoint of the line segment joining $\bm{c_i}$ and $\bm{c_j}$. Since $\bm{c_j}$ can be any other vertex, I check the distances for all $0 \le j \ne i < k$ and find the point of intersection $\bm{q_j}$ closest to $\bm{c_i}$. I denote that point as $\bm{q} = \bm{q_j}$, where $j \in [0, k)$ and $j \ne i$. Note that the point of intersection can be computed by solving a linear equation, which has a closed-form solution, and only those points must be considered that lie on the line segment joining $\bm{c_i}$ and $\bm{p}$. The closed-form solution makes it easy to compute partial derivatives.

\begin{figure}
\begin{center}
\begin{tikzpicture}
\def\thetaA{0}
\def\rA{1}
\def\thetaB{45}
\def\rB{1.25}
\def\thetaC{140}
\def\rC{1.1}
\def\thetaD{210}
\def\rD{1.3}
\def\thetaE{250}
\def\rE{1.4}
\def\thetaF{300}
\def\rF{1.3}

\def\alpha{1.4}

\def\thetaP{atan2(\yA + 0.6*(\yB - \yA), \xA + 0.6*(\xB - \xA))}
\def\rP{veclen(\xA + 0.6*(\xB - \xA), \yA + 0.6*(\yB - \yA))}
\def\rPD{\alpha*\rP}

\pgfmathsetmacro{\xA}{\rA*cos(\thetaA)}
\pgfmathsetmacro{\yA}{\rA*sin(\thetaA)}
\pgfmathsetmacro{\xB}{\rB*cos(\thetaB)}
\pgfmathsetmacro{\yB}{\rB*sin(\thetaB)}
\pgfmathsetmacro{\xC}{\rC*cos(\thetaC)}
\pgfmathsetmacro{\yC}{\rC*sin(\thetaC)}
\pgfmathsetmacro{\xD}{\rD*cos(\thetaD)}
\pgfmathsetmacro{\yD}{\rC*sin(\thetaD)}
\pgfmathsetmacro{\xE}{\rE*cos(\thetaE)}
\pgfmathsetmacro{\yE}{\rE*sin(\thetaE)}
\pgfmathsetmacro{\xF}{\rF*cos(\thetaF)}
\pgfmathsetmacro{\yF}{\rF*sin(\thetaF)}
\pgfmathsetmacro{\xP}{\rP*cos(\thetaP)}
\pgfmathsetmacro{\yP}{\rP*sin(\thetaP)}
\pgfmathsetmacro{\xPD}{\rPD*cos(\thetaP)}
\pgfmathsetmacro{\yPD}{\rPD*sin(\thetaP)}

\draw[thick] (\xA, \yA) -- (\xB, \yB) -- (\xC, \yC) -- (\xD, \yD) -- (\xE, \yE) -- (\xF, \yF) -- (\xA, \yA);

\draw[densely dash dot, thick] (\alpha*\xA, \alpha*\yA) -- (\alpha*\xB, \alpha*\yB) -- (\alpha*\xC, \alpha*\yC) -- (\alpha*\xD, \alpha*\yD) -- (\alpha*\xE, 1.3*\yE) -- (\alpha*\xF, \alpha*\yF) -- (\alpha*\xA, \alpha*\yA);

\draw[thin] (0,0) -- (\xP,\yP) -- (\xPD,\yPD);
\coordinate (A) at (0, 0);
\fill (A) circle (2pt) node[left] {$\bm{c_0}$};
\fill (\xP,\yP) circle (2pt) node[left] {$\bm{x}$};
\fill (\xPD,\yPD) circle (2pt) node[right] {$\bm{x'}$};
\end{tikzpicture}
\caption{Function blending: The vertex $\bm{c_0}$ is one of the seed vertices and the solid polygon around it is the boundary of the Voronoi cell of $\bm{c_0}$. Inside the cell, the relative weight of the linear function associated with the cell is $1$. The relative weight gradually reduces to $0$ as we move toward the dashed outer polygon. Outside the dashed outer polygon, the relative weight is $0$. For example, the weight is $1$ on the line segment $\bm{c_0x}$, and it linearly reduces to $0$ along the line segment $\bm{xx'}$. Note that $\left\|\bm{xx'}\right\| = \alpha_i \left\|\bm{c_0x}\right\|$, where $\alpha_i$ is the blending parameter.}
\label{fig:basis}
\end{center}
\end{figure}
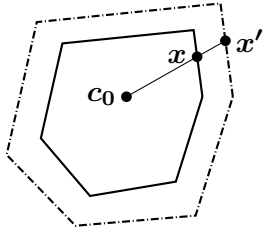

The ratio of the distance between $\bm{p}$ and $\bm{q}$ and the distance between $\bm{q_i}$ and $\bm{c_i}$ dictate the relative weight $\omega_i^{\mathrm{rel}}(\bm{p})$ of $L_i$ at $\bm{p}$. In my implementation, I used a linear relationship (w.r.t. the ratio) such that the relative weight vanishes when the ratio is $\alpha_i$, i.e.,
\begin{equation}
\omega_i^{\mathrm{rel}}(\bm{p}) = 1 - \frac{1}{\alpha_i}\frac{\left \|\bm{p} - \bm{q}\right \|}{\left \|\bm{c_i} - \bm{q}\right \|},
\label{eq:relative_weight}
\end{equation}
where $\left \|\cdot\right \|$ denotes the magnitude of a vector. If the distance ratio is greater than $\alpha_i$, the relative weight is $0$.
Note that if $\bm{p}$ is a point on the boundary of the Voronoi cell, $\omega_i^{\mathrm{rel}}(\bm{p}) = 1$. Fig.~\ref{fig:basis} shows how the relative weight varies in and around a Voronoi cell.

I compute the relative weights for the linear functions corresponding to all the cells and normalize them to compute the actual weight, i.e., 
\begin{equation}
\omega_i(\bm{p}) = \frac{\omega_i^{\mathrm{rel}}(\bm{p})}{\sum_{j = 0}^{k - 1} \omega_j^{\mathrm{rel}}(\bm{p})}.
\label{eq:normalized_weights}
\end{equation}
Note that the linear functions, $L_i$ for $0 \le i < k$, are continuous, and weight functions $\omega_i$ for $0 \le i < k$ are piecewise continuous over the whole domain. Therefore $\hat{f}$ is a piecewise-continuous function. Voronoi cell boundaries are continuous, but not differentiable, so it might be impossible to obtain a differentiable function with polygonal cells. If a differentiable or smooth function is desired, some future research on hyperspherical cells is needed.

\subsection{Objective Function}
In the previous section, I described how to compute the function given the parameters and some hyperparameters. In this section, I will describe the objective function that should be optimized to compute the parameters given the hyperparameters. Since the approximate function, $\hat{f}$, should be as close as possible to the input dataset, a natural objective function to minimize is
\begin{equation}
\sum_{i = 0}^{n - 1}(\hat{f}(\bm{a_i}) - b_i)^2
\label{eq:objective_function_no_regularization}
\end{equation}
over all parameters $\alpha_j$, $\bm{\beta_j}$, $\bm{c_j}$ for $0 \le j < k$, where $k$ is the number of cells and $\bm{a_i}$ and $b_i$ are input datasets as described in Section~\ref{subsection:input}. It can shown that if we assume a normal distribution of noise in the input dataset, minimizing the objective function above results in finding the most probable parameters.

In this paper, I focus on binary classification. For binary classification, the objective function that should be optimized is different. Here, $b_i \in \{0, 1\}$, i.e., each input data point either belongs to a class or it does not. Instead of using a discrete variable, the cellular learning algorithm computes the probability that a point $\bm{p}$ belongs to a certain class. That probability may be modeled as
\begin{equation}
\hat{F}(\bm{p}) = \frac{1}{1 + \exp(-\hat{f}(\bm{p}))},
\label{eq:probablity_binary}
\end{equation}
whose range is $(0, 1)$. It can be shown that in order to compute an approximate function $\hat{F}(\cdot)$, one should maximize the log-likelihood function
\begin{equation}
\hat{l} = \sum_{i = 0}^{n - 1} b_i \hat{f}(\bm{a_i}) - \log{\left(1 + \exp(\hat{f}(\bm{a_i}))\right)},
\label{eq:log_likelihood_no_regularization}
\end{equation}
where $\bm{a_i}$ and $b_i \in {0, 1}$ are the input datapoints as described in Section ~\ref{subsection:input} and $n$ is the the number of input datapoints. As in the case above, it can shown that if we assume a normal distribution of noise in the input dataset, maximizing the log-likelihood function above results in finding the most probable parameters.
\subsection{Regularization} \label{subsection:regularization}
If the objective functions above are optimized without any regularization terms, the resulting function will have a high variance, i.e., the function parameter will overfit the data. I will introduce two regularization terms that were successful in preventing high variance in our numerical experiments involving the MNIST dataset~\cite{deng2012mnist}.

First, we regularize the parameters in the linear functions corresponding to all the cells. I use the $L2$ regularization,
\begin{equation}
R_{\bm{\beta}} = \sum_{i = 0}^{k - 1} \sum_{j = 0}^d \beta_{ij}^2 = \sum_{i = 0}^{k - 1} \left\|\bm{\beta_i}\right\|_2^2,
\label{eq:beta_l2_regularizartion}
\end{equation}
where $k$ is the number of cells, $d$ is the number of dimensions, and $\beta_{ij}$ are coefficients of the terms in linear functions as described in ~\ref{subsection:parameters}. This is a standard regularization used in many machine learning algorithms. It ensures that none of the coefficients are too large. From a probabilistic perspective, $L2$ regularization assumes a normal distribution for the prior probability of the value of the coefficients. 

Second, we regularize the blending parameters (the $\alpha$ parameters). If the blending parameters for all cells are all $0$ (or close to it), there is no practical blending of the linear functions, and the resulting function overfits the training data (observed in my initial numerical experiments without this regularization). In order to ensure the blending parameters are greater than $0$, we simply use their reciprocal for regularization, i.e., 
\begin{equation}
R_{\alpha} = \sum_{i = 0}^{k - 1} \frac{1}{\alpha_i},
\label{eq:alpha_regularizartion}
\end{equation}
where $k$ is the number of cells and $\alpha_i$ is the blending parameter associated with cell $i$. This regularization may be viewed as an $L1$ regularization of the reciprocal of the blending parameters. From a probabilistic perspective, $L1$ regularization assumes a Laplace distribution for the prior probability of the reciprocal of the value of the blending parameters.

In my numerical experiments with the MNIST dataset, I did not need to regularize the parameters associated with the location of the cell vertices. For other applications, one can consider regularization for these parameters, too.

With the inclusion of the regularization terms, the objective function that should be maximized for binary classification is
\begin{equation}
\hat{l} = \sum_{i = 0}^{n - 1} \left(b_i \hat{f}(\bm{a_i}) - \log{\left(1 + \exp\left(\hat{f}(\bm{a_i})\right)\right)}\right) - \lambda_{\alpha} \sum_{i = 0}^{k - 1} \frac{1}{\alpha_i} - \lambda_{\beta} \sum_{i = 0}^{k - 1} \left\|\bm{\beta_i}\right\|_2^2,
\label{eq:final_objective_function_with_regularizartion}
\end{equation}
where $\lambda_{\alpha}$ and $\lambda_{\beta}$ are new hyperparameters that control the extent of regularization. 
\subsection{Algorithm}
In the previous sections, I explained the rationale behind the objective function that should be maximized for binary classification. Now we are ready to use the objective function and develop the cellular learning algorithm. Algorithm~\ref{algorithm:main} shows the algorithm. Note that the Adam optimization step is the slowest step in the algorithm. For computational efficiency, an implementation of the Adam optimization algorithm should compute analytical partial derivatives of the objective function with respect to all the parameters.

\begin{algorithm}[H]
\SetAlgoLined
\KwData{$n$, $\bm{a_i}$, and $b_i$ for $0 \le i < n$}
\KwIn{Hyperparameters: $k$, $\lambda_{\alpha}$, $\lambda_{\bm{\beta}}$}
\KwResult{Parameters: $\alpha_i$, $\bm{\beta_i}$, and $\bm{c_i}$ for $0 \le i < k$}
Initialize $\alpha_i$ to some value, say, $0.3$.\\
Initialize $\bm{\beta_i} = 0$.\\
Randomly pick $k$ seed vertices from the input dataset and initialize all $\bm{c_i}$.\\
Run the Lloyd's $k$-means clustering algorithm to find initial cluster centers $\bm{c_i}$, where $0 \le i < k$.\\
Use the Adam optimization algorithm to maximize the objective function.\\
Return the parameters obtained above.
\caption{The Cellular Learning Algorithm}
\label{algorithm:main}
\end{algorithm}

\subsection{Limitations}
In the current implementation, the complexity of computing the Voronoi cell boundaries is $\Theta(n^2)$, where $n$ is the number of cells. It can improved to $\Theta(\log{n})$ by considering a hierarchical network (see Section~\ref{section:future}), but if the blending parameters are large, the complexity may still be $\Theta(n^2)$ because every cell may influence many other cells. With Voronoi cells, it is impossible to obtain a differentiable function because Voronoi cells themselves have sharp corners. Unlike neural networks, cellular networks do not exploit any periodicity in the data. 

\section{Experiment}
I implemented the algorithm described above in C++ and parallelized it using the OpenMP framework. For the Adam optimization algorithm, I set the parameters to the recommended value in the paper~\cite{kingma14}. For the stochastic algorithm, I used roughly 5\% of data points for every iteration (the minibatch size). In the code for computing the partial derivatives, at points where the derivative is not continuous, I arbitrarily decided the ``piece'' of the function for which I computed the derivative. Since the likelihood of having to compute the derivative at such locations is small, the choice of the subderivative should not affect the optimization process much. 

\subsection{MNIST Dataset}
I used the MNIST dataset as the input to test my algorithm. It is a set of 60,000 greyscale images of handwritten digits from 0 to 9 that fit into a 28x28 pixel bounding box. The 60,000 images are used to train the model. There are 10,000 additional images for testing the model. Since the MNIST dataset is a standard dataset that has been used to test and validate many machine learning algorithms, I have also used it to test my algorithm.

\subsection{One v. Rest}
With the MNIST dataset, the task is to classify an image into $1$ of $10$ digits (from $0$ to $9$), but the objective function at the end of Section~\ref{subsection:regularization} can classify a data point into one of two classifications. In order to use the objective function for the MNIST dataset, we have to use the one-v-rest (OvR) strategy. We have to train $10$ classifiers. The first classifier is trained to distinguish between $0$ and the rest of the digits, i.e., it returns the probability of an image being $0$, and the second classifier is trained to distinguish between $1$ and the rest of the digits, i.e., it returns the probability of an image being $1$, and so on. Finally, the image is classified as the digit with the highest probability.

\subsection{Results}
I ran my implementation on the MNIST dataset and recorded the accuracy for a set of hyperparameters. In this section, I will report the results of my numerical experiments. 

In my first set of experiments, I used $k = 30$, $k = 40$, and $k = 50$ cells with the seed vertices distributed among the $60,000$ data points in the training set. I used the same initial seed vertex distribution for all $10$ classifiers in the OvR strategy. I found that the results improved when we increased the number of cells from $30$ to $40$, but there was no significant improvement when I increased the number of cells from $40$ to $50$. Therefore, I will report the results only for $k = 40$ cells.

In my second set of experiments, I distributed seed vertices differently for each classifier in the OvR strategy. If the binary classifier was classifying an image between digit $i$ and other digits, I used $10$ seed vertices distributed among training images for digit $i$ and $4$ seed vertices for each of the other digits, which results in $10 + 4 * 9 = 46$ seed vertices.

For both sets of experiments, I ran the Adam optimization algorithm for $30$ and $60$ epochs. The accuracies I obtained from the experiments are shown in Tables~\ref{table:40_30},~\ref{table:40_60},~\ref{table:10_4_30}, and~\ref{table:10_4_60}. Without any regularization, I obtained an accuracy of only $97.25\%$ on the testing set. With regularization, the best accuracy of $98.20\%$ was from using $46$ cells after $60$ epochs. 

Note that the total number of parameters for each cellular network is $2k(d + 1)$ (see Section~\ref{subsection:parameters}), where k is the number of cells and $d$ is the number of dimensions. There are $10$ networks (one for each digit in the OvR strategy). For $40$ cells, we compute $628,000$ parameters, and for $46$ cells, we compute $722,200$ parameters.

The training takes around $3$ hours on $16$ cores of a shared memory processor on Google Cloud's \emph{e2-standard-16} machines. Since floating point operations are not associative, there is an uncertainty of around 0.05\% in the results, which was estimated after repeated running of the code with the same hyperparameters.

\begin{table}
\centering
\begin{tabular}{|c|c|c|c|c|c|c|}
\hline
\multirow{2}{*}{$\lambda_{\alpha}$} & \multicolumn{6}{c|}{$\lambda_{\beta}$} \\
\cline{2-7}
& 0.0001 & 0.00001 & 0.0000005 & 0.0000001 & 0.00000005 & 0.00000001\\
\hline
0.1250 & 97.92\% & 97.85\% & 97.87\% & 97.87\% & 97.87\% & 97.92\% \\
0.0950 & 98.00\% & 97.89\% & 97.91\% & 97.97\% & 97.99\% & 97.99\% \\
0.0625 & 97.89\% & 97.92\% & 97.95\% & 97.94\% & 97.91\% & 97.89\% \\
0.0450 & 97.93\% & 97.92\% & \textbf{98.02\%} & 97.94\% & 97.90\% & 97.87\% \\
0.0300 & 97.93\% & 97.92\% & 97.85\% & 97.83\% & 97.78\% & 97.88\% \\
0.0150 & 97.86\% & 97.77\% & 97.86\% & 97.87\% & 97.71\% & 97.84\% \\
\hline
\end{tabular}
\caption{Accuracy of the cellular learning algorithm with 40 cells in the network after 30 epochs.}
\label{table:40_30}
\end{table}

\begin{table}
\centering
\begin{tabular}{|c|c|c|c|c|c|}
\hline
\multirow{2}{*}{$\lambda_{\alpha}$} & \multicolumn{5}{c|}{$\lambda_{\beta}$} \\
\cline{2-6}
& 0.00005 & 0.00001 & 0.000005 & 0.000001 & 0.0000005\\
\hline
0.250 & 97.89\% & 97.87\% & 97.91\% & 97.94\% & 97.95\% \\
0.100 & 97.96\% & 98.02\% & 98.00\% & 98.14\% & 97.98\% \\
0.075 & 98.02\% & 98.02\% & 97.95\% & 98.14\% & 97.93\% \\
0.065 & 98.01\% & 97.98\% & \textbf{98.12\%} & 98.09\% & 98.10\% \\
0.050 & 98.02\% & 98.07\% & 98.09\% & 98.09\% & 98.04\% \\
0.025 & 97.92\% & 98.04\% & 98.03\% & 97.91\% & 98.03\% \\
\hline
\end{tabular}
\caption{Accuracy of the cellular learning algorithm with 40 cells in the network after 60 epochs.}
\label{table:40_60}
\end{table}

\begin{table}
\centering
\begin{tabular}{|c|c|c|c|c|c|}
\hline
\multirow{2}{*}{$\lambda_{\alpha}$} & \multicolumn{5}{c|}{$\lambda_{\beta}$} \\
\cline{2-6}
& 0.0005 & 0.0001 & 0.00005 & 0.00001 & 0.000005\\
\hline
0.250 & 97.88\% & 97.88\% & 97.86\% & 97.90\% & 97.91\% \\
0.100 & 97.87\% & 97.90\% & 97.90\% & 97.93\% & 97.94\% \\
0.075 & 97.93\% & 97.89\% & 97.88\% & 97.93\% & 97.86\% \\
0.050 & 97.95\% & 97.92\% & 97.95\% & 97.92\% & 97.85\% \\
0.025 & 97.93\% & 97.85\% & \textbf{97.97\%} & 97.91\% & 97.93\% \\
0.010 & 97.71\% & 97.81\% & 97.84\% & 97.80\% & 97.78\% \\
\hline
\end{tabular}
\caption{Accuracy of the cellular learning algorithm with 46 cells in the network after 30 epochs. In the OvR strategy, one of the digits has 10 cells and the other digits have 4 cells each.}
\label{table:10_4_30}
\end{table}

\begin{table}
\centering
\begin{tabular}{|c|c|c|c|c|c|}
\hline
\multirow{2}{*}{$\lambda_{\alpha}$} & \multicolumn{5}{c|}{$\lambda_{\beta}$} \\
\cline{2-6}
& 0.005 & 0.0025 & 0.001 & 0.00075 & 0.0005\\
\hline
0.100 & 98.02\% & 98.01\% & 98.12\% & 98.00\% & 98.01\% \\
0.085 & 98.01\% & 98.01\% & 97.94\% & 98.11\% & 97.97\% \\
0.075 & 98.04\% & 98.06\% & \textbf{98.20\%} & 98.05\% & 98.01\% \\
0.065 & 97.89\% & 98.08\% & 98.04\% & 98.10\% & 98.03\% \\
0.050 & 97.97\% & 97.96\% & 98.08\% & 97.98\% & 97.92\% \\
0.025 & 97.93\% & 98.02\% & 97.96\% & 98.02\% & 97.97\% \\
\hline
\end{tabular}
\caption{Accuracy of the cellular learning algorithm with 46 cells in the network after 60 epochs. In the OvR strategy, one of the digits has 10 cells and the other digits have 4 cells each.}
\label{table:10_4_60}
\end{table}

\section{Future Work} \label{section:future}
There are multiple avenues for future research. In my algorithm, I have used a linear function within a cell. Instead, one can consider a quadratic or some other nonlinear function. Further, one can consider using yet another cellular network in place of a linear function. This would result in a hierarchical cellular network. With such a network, the complexity of computing the function can be reduced to $\Theta(\log{n})$ from $\Theta(n^2)$, where $n$ is the number of cells in the network. A hierarchical approach results in a scalable algorithm that can compute highly nonlinear functions quickly. In order to ensure the $\Theta(\log{n})$ complexity, the number of cells in every cell of the hierarchical network should be bounded from above by a constant, and the blending parameters should have a low value. 

Additionally, approaches involving smart initial seeding and adaptive refinement of vertices should help us quickly arrive at a function approximating our scattered data. Many adaptive refinement techniques have been used successfully for solving partial difference equations over unstructured meshes. Such techniques should also be helpful here.

I have used polygonal cells (Voronoi cells are polygonal) that are implicitly constructed by seed vertices. Instead of polygonal cells, one can consider hyperspherical cells. With  hyperspherical cells, there will be regions that are not under the influence of any cell, but the boundary of hyperspherical cells is easier to compute. There may be applications where such cells are appropriate. This approach is similar to RBF networks with a compact basis function. When combined with hierarchical networks, this approach may be fast and scalable. A sophisticated combination of the above avenues can lead to fast, robust, scalable, and explainable algorithms for regression in high dimensions.

\pagebreak
\bibliographystyle{abbrv}
\bibliography{refs}

\begin{thebibliography}{10}

\bibitem{bishop05}
C.~M. Bishop.
\newblock {\em Pattern Recognition and Machine Learning (Information Science
  and Statistics)}.
\newblock Springer-Verlag, Berlin, Heidelberg, 2006.

\bibitem{Broomhead1988b}
D.~S. Broomhead and D.~Lowe.
\newblock Multivariable functional interpolation and adaptive networks.
\newblock {\em Complex Syst.}, 2, 1988.

\bibitem{Broomhead1988a}
D.~S. Broomhead and D.~Lowe.
\newblock Radial basis functions, multi-variable functional interpolation and
  adaptive networks.
\newblock 1988.

\bibitem{chen22}
K.-L. Chen, H.~Garudadri, and B.~D. Rao.
\newblock Improved bounds on neural complexity for representing piecewise
  linear functions.
\newblock In {\em Proceedings of the 36th International Conference on Neural
  Information Processing Systems}, NIPS '22, Red Hook, NY, USA, 2022. Curran
  Associates Inc.

\bibitem{cortes-vapnik-ml95}
C.~Cortes and V.~Vapnik.
\newblock Support-vector networks.
\newblock {\em Machine Learning}, 20(3):273--297, 1995.

\bibitem{deng2012mnist}
L.~Deng.
\newblock The mnist database of handwritten digit images for machine learning
  research.
\newblock {\em IEEE Signal Processing Magazine}, 29(6):141--142, 2012.

\bibitem{fukushima80}
K.~Fukushima.
\newblock Neocognitron: A self-organizing neural network model for a mechanism
  of pattern recognition unaffected by shift in position.
\newblock {\em Biological Cybernetics}, 36:193--202, 1980.

\bibitem{kingma14}
D.~P. Kingma and J.~Ba.
\newblock Adam: A method for stochastic optimization.
\newblock {\em CoRR}, abs/1412.6980, 2014.

\bibitem{lloyd82}
S.~Lloyd.
\newblock Least squares quantization in pcm.
\newblock {\em IEEE Transactions on Information Theory}, 28(2):129--137, 1982.

\bibitem{murphy13}
K.~P. Murphy.
\newblock {\em Machine learning : a probabilistic perspective}.
\newblock MIT Press, Cambridge, Mass. [u.a.], 2013.

\bibitem{polianskii19}
V.~Polianskii and F.~T. Pokorny.
\newblock Voronoi boundary classification: A high-dimensional geometric
  approach via weighted {M}onte {C}arlo integration.
\newblock In K.~Chaudhuri and R.~Salakhutdinov, editors, {\em Proceedings of
  the 36th International Conference on Machine Learning}, volume~97 of {\em
  Proceedings of Machine Learning Research}, pages 5162--5170. PMLR, 09--15 Jun
  2019.

\bibitem{polianskii20}
V.~Polianskii and F.~T. Pokorny.
\newblock Voronoi graph traversal in high dimensions with applications to
  topological data analysis and piecewise linear interpolation.
\newblock In {\em Proceedings of the 26th ACM SIGKDD International Conference
  on Knowledge Discovery \& Data Mining}, KDD '20, page 2154–2164, New York,
  NY, USA, 2020. Association for Computing Machinery.

\bibitem{siahkamari20}
A.~Siahkamari, A.~Gangrade, B.~Kulis, and V.~Saligrama.
\newblock Piecewise linear regression via a difference of convex functions.
\newblock In H.~D. III and A.~Singh, editors, {\em Proceedings of the 37th
  International Conference on Machine Learning}, volume 119 of {\em Proceedings
  of Machine Learning Research}, pages 8895--8904. PMLR, 13--18 Jul 2020.

\bibitem{toriello12}
A.~Toriello and J.~P. Vielma.
\newblock Fitting piecewise linear continuous functions.
\newblock {\em European Journal of Operational Research}, 219(1):86--95, 2012.

\bibitem{tsang05}
I.~W. Tsang, J.~T. Kwok, and P.-M. Cheung.
\newblock Core vector machines: Fast svm training on very large data sets.
\newblock {\em Journal of Machine Learning Research}, 6(13):363--392, 2005.

\bibitem{vanbelle2016}
V.~Van~Belle, B.~Van~Calster, S.~Van~Huffel, J.~Suykens, and P.~Lisboa.
\newblock Explaining support vector machines: A color based nomogram.
\newblock {\em PLOS ONE}, 11(10):e0164568, 2016.

\bibitem{voronoi1908a}
G.~Voronoi.
\newblock Nouvelles applications des param{\`e}tres continus {\`a} la
  th{\'e}orie des formes quadratiques. premier m{\'e}moire. sur quelques
  propri{\'e}t{\'e}s des formes quadratiques positives parfaites.
\newblock {\em Journal f{\"u}r die reine und angewandte Mathematik (Crelles
  Journal)}, 1908:97 -- 102, 1908.

\bibitem{wang17}
H.~Wang, J.~Xiong, Z.~Yao, M.~Lin, and J.~Ren.
\newblock Research survey on support vector machine.
\newblock MOBIMEDIA'17, page 95–103, Brussels, BEL, 2017. ICST (Institute for
  Computer Sciences, Social-Informatics and Telecommunications Engineering).

\end{thebibliography}
\end{document}